\documentstyle[11pt]{article}
\begin{document}

\title{\textbf{The effect of density stratification on the resonant absorption of MHD waves in coronal loops}}
\author{K. Karami$^{1,2}$\thanks{E-mail: KKarami@uok.ac.ir} , S. Nasiri$^{3,4}$, S.
Amiri$^{4}$\\\\
$^{1}$\small{Department of Physics, University of Kurdistan,
Pasdaran St., Sanandaj, Iran}\\$^{2}$\small{Research Institute for
Astronomy $\&$ Astrophysics of Maragha (RIAAM), Maragha, Iran}\\
$^{3}$\small{Institute for Advanced Studies in Basic Sciences
(IASBS), Gava Zang, Zanjan, Iran}\\
$^{4}$\small{Department of Physics, Zanjan University, Zanjan,
Iran}} \maketitle

\begin{abstract}
The standing quasi-modes of the ideal MHD in a zero-$\beta$
cylindrical magnetic flux tube that undergoes a longitudinal
density stratification and radial density structuring is
considered. The radial structuring is assumed to be a linearly
varying density profile. Using the relevant connection formulae of
the resonant absorption, the dispersion relation for the fast MHD
body waves is derived and solved numerically to obtain both the
frequencies and damping rates of the fundamental and
first-overtone, $k=1,2$ modes of both the kink ($m=1$), and
fluting ($m=2$) waves. Where $k$ and $m$ are the longitudinal and
azimuthal mode numbers, respectively.
\end{abstract}

\noindent{\textbf{Key words:}~~~Sun: corona -- Sun: magnetic
fields -- Sun: oscillations}
%-----------------------------------------------------------------------------------------------
\clearpage
\section{Introduction}
Transverse oscillations of coronal loops were first identified by
Aschwanden et al. (1999) and Nakariakov et al. (1999) using the
observations of TRACE. Nakariakov et al. (1999) reported the
detection of spatial oscillations in five coronal loops with
periods ranging from 258 to 320 s. The decay time was
$(14.5\pm2.7)$ minutes for an oscillation of $(3.9\pm0.13)$
millihertz. Also Wang \& Solanki (2004) described a loop
oscillation observed on 17 April 2002 by TRACE in 195${\AA}$. They
interpreted the observed loop motion as a vertical oscillation,
with a period of 3.9 minutes and a decay time of 11.9 minutes. All
these observations indicate strong dissipation of the wave energy
that may be the cause of coronal heating.

Since the discovery of the hot solar corona about 66 years ago,
different theories of coronal heating have been put forward and
debated. Ionson (1978) was first to suggest that the resonant
absorption of MHD waves in coronal plasmas could be a primary
mechanism in coronal heating. Since then, much analytical and
numerical work has been done on the subject. Rae and Roberts
(1982) investigated both eikonal and differential equation
approaches for the propagation of MHD waves in inhomogeneous
plasmas. Hollweg (1987a,b) considered a dissipative layer
 of planar geometry to study the resonant absorption of coronal loops.
 Poedts et al. (1989, 1990) developed a finite element code to elaborate on the resonant absorption of Alfv\'{e}n
waves in circular cylinders.

 Davila (1987) and Steinolfson \& Davila (1993) did much work on resonant absorption through
resistivity. Goossens et al. (2002) used the TRACE data of Ofman
\& Aschwanden (2002) to infer the width of the inhomogeneous layer
for 11 coronal loops. Ruderman \& Roberts (2002) did similar
analysis with the data of Nakariakov et al. (1999). Van
Doorsselaere et al. (2004a) used the LEDA code to study the
resistive absorption of the kink modes of cylindrical models. They
concluded that, when the width of the nonuniform layer was
increased, their numerical results differed  by as much as 25$\%$
from those obtained with the analytical approximation. In the
vicinity of singularity, field gradients are large. Recognizing
this, Sakurai et al. (1991a,b) and Goossens et al. (1992, 1995)
developed a method to analyze dissipative processes in such
regimes and to neglect them elsewhere.

Safari et al. (2006) studied the resonant absorption of MHD waves
in magnetized flux tubes with a radial density inhomogeneity.
Using the approximation that resistive and viscous processes are
operative in thin layers surrounding the singularities of the MHD
equations, they obtained the full spectrum of the eigenfrequencies
and damping rates of the quasi-modes of the ideal MHD of the tube.
Both surface and body modes were analyzed.

Verwichte et al. (2004), using the observations of TRACE, detected
the multimode oscillations for the first time. They found that two
loops are oscillating in both the fundamental and the
first-overtone standing fast kink modes. According to the theory
of MHD waves, for uniform loops the ratio of the period of the
fundamental to the period of the first overtone is exactly 2, but
the ratios found by Verwichte et al. (2004) are 1.81 and 1.64.
However, these values were corrected to 1.82 and 1.58,
respectively, by Van Doorsselaere et al. (2007) and thus clearly
differ from 2. This may be caused by different factors such as the
effects of curvature (see e.g. Van Doorsselaere et al. 2004b),
leakage (see De Pontieu et al. 2001), magnetic twist (see e.g.
Erd\'{e}lyi \& Fedun 2006; Erd\'{e}lyi \& Carter 2006) and density
stratification in the loops (see Andries et al. 2005b; Erdelyi \&
Verth 2007). However, now it clear that the only cause of this
deviation is the longitudinal density stratification. Curvature
does not affect sufficiently the frequencies. Leakage can cause
the damping of oscillations but practically does not affect the
frequencies. Recently Ruderman (2007) showed that the magnetic
twist also does not affect the frequencies.

Andries et al. (2005b) considered a coronal loop model with a
straight cylindrically magnetized flux tube. They elaborated the
effect of longitudinal density stratification on the oscillation
frequencies and the damping rates of fast surface waves by
resonant absorption.

Andries et al. (2005a), with the help of the method introduced by
Andries et al. (2005b), showed that in the presence of a
longitudinal density stratification the ratio of the periods of
the fundamental kink mode of a coronal loop and of its first
harmonic is lower than 2. They used this ratio to estimate the
density scaleheight in the solar atmosphere.

Arregui et al. (2005) studied the effects of both radial and
longitudinal density stratifications on the frequency and damping
rate of resonantly fast kink modes. Using a numerical code
(POLLUX), they solved the linear resistive MHD equations for a
cylindrical flux tube model with a wide range of values of several
loop parameters.

Safari et al. (2007) investigated the oscillations of coronal
loops in presence of vertical stratification for a zero-$\beta$
plasma. They solved the radial equation in the thin tube
approximation and the transverse equation by both perturbational
and numerical techniques. They confirmed the result obtained by
Andries et al. (2005a) that the ratio of periods of the
fundamental and first-overtone modes differ from 2 in loops with
longitudinal density stratification.

Karami \& Asvar (2007, hereafter Paper I) studied both the
oscillations and damping of standing fast MHD body waves in
coronal loops in the presence of both longitudinal density
stratification and radial density structuring. The radial
structuring was assumed to have a step-like density profile. They
derived both the frequencies and damping rates of the fundamental,
first-overtone and second-overtone frequencies of both the kink
and fluting modes. They obtained that the ratios of the
frequencies of the first-overtone and its fundamental mode for
both the kink and fluting modes are lower than 2 (for unstratified
loops).

In the present work, our aim is to investigate the effects of
longitudinal density stratification and radial density structuring
on resonant absorption of standing quasi-modes of the ideal MHD in
the cold coronal loops observed by Verwichte et al. (2004) deduced
from the TRACE data. This paper is organized as follows. In
Sections 2 and 3 we combine the two techniques of Paper I and of
Thompson \& Wright (1993) to derive the equations of motion,
introduce the relevant connection formulae and obtain the
dispersion relation. In Section 4 we give numerical results.
Section 5 is devoted to conclusions.

%-----------------------------------------------------------------------------------------------
\section{Equations of motion}

The linearized MHD equations for a zero-$\beta$ plasma are
\begin{eqnarray} \frac{\partial\delta\mathbf{v}}{\partial
t}=\frac{1}{4\pi\rho}\{(\nabla\times\delta\mathbf{B})\times\mathbf{B}
+(\nabla\times\mathbf{B})\times\delta\mathbf{B}\}
+\frac{\eta}{\rho}\nabla^2\delta\mathbf{v},\label{mhd1}
\end{eqnarray}
\begin{eqnarray}
\frac{\partial\delta\mathbf{B}}{\partial
t}=\nabla\times(\delta\mathbf{v}\times\mathbf{B})+
\frac{c^2}{4\pi\sigma}\nabla^2\delta\mathbf{B},\label{mhd2}
\end{eqnarray}
where $\delta\bf{v}$ and $\delta\bf{B}$ are the Eulerian
perturbations in the velocity and magnetic fields; $\rho$,
$\sigma$, $\eta$ and $c$ are the mass density, the electrical
conductivity, the viscosity and the speed of light, respectively.
The simplifying assumptions are the same as Paper I.

From Andries et al. (2005b), in the absence of dissipations, the
perturbed quantities $\delta{\bf{B}}(r,z)$ and
$\delta{\bf{v}}(r,z)$ can be expressed in terms of a complete set
of orthonormal functions $\psi^{(k)}(z)$ as follows
\begin{eqnarray}
\delta{\mathbf{B}}(r,z)=\sum_{k=1}^{\infty}\delta{\mathbf{B}}^{(k)}(r)\psi^{(k)}(z),\label{Brz}
\end{eqnarray}
where the same relation is kept for $\delta{\bf{v}}(r,z)$. Note
that $\psi^{(k)}(z)$ itself satisfies the eigenvalue relation
$L_A\psi^{(k)}(z)=\lambda_k\psi^{(k)}(z)$, where $L_{\rm A}$ is
the Alfv\'{e}n operator,
\begin{eqnarray}
L_A=\rho\omega^2+\frac{B^2}{4\pi}\frac{\partial^2}{\partial
z^2}=\rho\Big(\omega^2+v_A^2\frac{\partial^2}{\partial z^2}\Big),
\end{eqnarray}
with Alfv\'{e}n velocity $v_A=\frac{B}{\sqrt{4\pi\rho}}$ and
straight constant background magnetic filed ${\bf B}=B\hat{{\bf
z}}$. One may show that the resonance absorption occurs where the
Alfv\'{e}n operator has a vanishing eigenvalue.

Let us denote the radius of the tube by $R$ and a radius at with
the resonant absorption occurs by $R_1<R$. The thickness of the
inhomogeneous layer, $l=R-R_1$, will be assumed to be small. From
Safari et al. (2006) and Paper I, the density profile is assumed
to be
\begin{eqnarray}
\rho(r,z)&=&\rho_0(r)\exp{\Big[-\mu\sin\big(\frac{\pi
z}{L}\big)\Big]},~~~\mu:=\frac{L}{\pi H},\nonumber\\
\rho_0(r)&=&\left\{\begin{array}{ccc}
    \rho_{{\rm in}},&(r<R_1),&\\
    \Big[\frac{\rho_{\rm in}-\rho_{\rm ex}}{R-R_1}\Big](R-r)+\rho_{\rm ex},&(R_1<r<R),&\\
      \rho_{{\rm ex}},&(r>R),&\\
      \end{array}\right.
\end{eqnarray}
where $\mu$ is defined as stratification parameter, $H$ and $L$
are the density scale hight and length of the loop, respectively.
Also $\rho_{{\rm in}}$ and $\rho_{{\rm ex}}$ are the interior and
exterior constant densities of the tube, respectively.

From paper I, in the absence of dissipations, in the interior
region, $r<R_1$, solutions of Eqs. (\ref{mhd1})-(\ref{mhd2}) are
\begin{equation} \delta
B_z^{({\rm in},k)}(r)=\left\{
\begin{array}{cccc} I_m(|k_{{\rm in},k}|r),&k_{{\rm in},k}^2<0,&{\rm surface~waves},&\\
J_m(|k_{{\rm in},k}|r),&k_{{\rm in},k}^2>0,&{\rm
body~waves},&\\k_{{\rm in},k}^2=\frac{\lambda_{{\rm
in},k}}{B^2/4\pi},&\end{array}\right.\label{Bzrin}
\end{equation}
where $J_m$ and $I_m$ are Bessel and modified Bessel functions of
the first kind, respectively.
 In the exterior region, $r>R$, the waves should be evanescent.
The solutions are
\begin{eqnarray}
\delta B_z^{({\rm ex},k)}(r)=K_m(k_{{\rm ex},k}r),~~~~~~k_{{\rm
ex},k}^2=-\frac{\lambda_{{\rm ex},k}}{B^2/4\pi}>0,\label{Bzrex}
 \end{eqnarray}
where $K_m$ is the modified Bessel function of the second kind.

Here we only consider the body waves. Because under coronal
conditions, $v_{\rm A_{in}}< v_{\rm A_{ex}}$, magnetic flux tube
supports fast body oscillations and there are no longer any
surface modes (see Fig. 4 in Edwin \& Roberts 1983). From Eq. (7)
in Paper I and Eqs. (\ref{Brz}), (\ref{Bzrin}), one can obtain
$\delta B_z^{({\rm in},k)}(r,z)$ and $\delta v_r^{({\rm
in},k)}(r,z)$ in the interior region $(r<R_1)$ as
\begin{eqnarray}
\delta B_z^{({\rm in})}(r,z)&=&\sum_{k=1}^{+\infty}A^{({\rm
in},k)}J_m(|k_{{\rm in},k}|r)\psi^{({\rm in},k)}(z), \nonumber\\
\delta v_r^{({\rm in})}(r,z)&=&-\frac{i\omega
B}{4\pi}\sum_{k=1}^{+\infty}\frac{k_{{\rm in},k}}{\lambda_{{\rm
in},k}}A^{({\rm in},k)}J'_m(|k_{{\rm in},k}|r)\psi^{({\rm
in},k)}(z),\nonumber\\\label{BzVrrz}
\end{eqnarray}
where a prime on $J_m$ and hereafter on each function indicates a
derivative with respect to their appropriate arguments. The
results for the exterior region are the same as Eq.
(\ref{BzVrrz}), except that $J_m$ and index ''in'' are replaced by
$K_m$ and ''ex'' everywhere.
%-----------------------------------------------------------------------------------------------
\section{Connection formulae, dispersion relation and damping}
From Andries et al. (2005b) and according to the connection
formulae developed by Thompson \& Wright (1993), the jump across
the boundary for $\delta B_z$ and $\delta v_r$ are
 \begin{eqnarray}
[\delta B_z]=0,\label{Bjump}
\end{eqnarray}

\begin{eqnarray}
[\delta v_r]=-\sum_{k=1}^{+\infty}\frac{B\tilde{\omega}
m^2\Big<\phi^{({\rm in},k)}\Big|\delta B_z^{({\rm
in},k)}\Big>}{4r_A^2\Big<\phi^{({\rm
in},k)}\Big|L_{A_1}\Big|\phi^{({\rm in},k)}\Big>}\phi^{({\rm
in},k)},\label{vjump}
\end{eqnarray}
where
\begin{eqnarray}
\phi^{({\rm
in},k)}=\sqrt{\frac{2}{L}}\sum_{j=1}^{+\infty}\phi_j^{({\rm
in},k)}\sin\left(\frac{j\pi}{L}z\right),
\end{eqnarray}
\begin{eqnarray}
L_A\phi^{({\rm in},k)}=0~\Longrightarrow~\phi_j^{({\rm in},k)}
=\Big\{ {\begin{array}{*{20}c}
   \frac{k^2S_{kj}}{\rho_{\rm in}(1+S_{kk})(j^2-k^2)} & {k\ne j}  \\
   1 & {k=j}  \\
\end{array}},
\end{eqnarray}
and
\begin{eqnarray}
L_{A1}=\frac{\partial{L_A}}{\partial r}\Big
{|}_{r=r_A}=\tilde{\omega}^2
(1+S_{kk})\frac{\partial\rho_{0}(r)}{\partial
 r}\Big{|}_{r=r_A},
 \end{eqnarray}
\begin{eqnarray}
S_{kj}=\frac{2}{L}\int_0^L\sin(\frac{k\pi}{L} z)\Big
[-\mu\sin(\frac{\pi}{L} z)\Big
 ]\sin(\frac{j\pi}{L} z)dz,
 \end{eqnarray}
\begin{eqnarray}
\Big<\phi^{({\rm in},k)}\Big|L_{A_1}\Big|\phi^{({\rm
in},k)}\Big>=\frac{\rho _{\rm ex}-\rho _{\rm
in}}{R-R_1}\tilde{\omega}^2(1 + S_{kk}).
\end{eqnarray}
Note that $R_1<r_A<R$~ is the radius at which the singularity
occurs and, $\tilde{\omega}=\omega+i\gamma$ where $\gamma$ is
damping rate. Substituting the
 fields of Eq. (\ref{BzVrrz}) in jump conditions, gives
\begin{eqnarray}
 \left( {\begin{array}{*{20}c}
   {\Pi_1^{(\rm ex,1)} } & { - \Pi_1^{(\rm in,1)} } & {\Pi_1^{(\rm ex,2)} } & { - \Pi_1^{(\rm in,2)} } &  \ldots   \\
   {\Xi_1^{(\rm ex,1)} } & {\Xi_1^{(\rm in,1)}  + {\mathcal{D}}_1^{(\rm in,1)} } & {\Xi_1^{(\rm ex,2)} } & {\Xi_1^{(\rm in,2)}  + {\mathcal{D}}_1^{(\rm in,2)} } &  \ldots   \\
   {\Pi_2^{(\rm ex,1)} } & { - \Pi_2^{(\rm in,1)} } & {\Pi_2^{(\rm ex,2)} } & { - \Pi_2^{(\rm in,2)} } &  \ldots   \\
   {\Xi_2^{(\rm ex,1)} } & {\Xi_2^{(\rm in,1)}  + {\mathcal{D}}_2^{(\rm in,1)} } & {\Xi_2^{(\rm ex,2)} } & {\Xi_2^{(\rm in,2)}  + {\mathcal{D}}_2^{(\rm in,2)} } &  \ldots   \\
    \vdots  &  \vdots  &  \vdots  &  \vdots  &  \ddots   \\
\end{array}}\right)
\left( {\begin{array}{*{20}c}
   {A^{(\rm ex,1)} }  \\
   {A^{(\rm in,1)} }  \\
   {A^{(\rm ex,2)} }  \\
   {A^{(\rm in,2)} }  \\
    \vdots   \\
\end{array}} \right)
= 0,\label{dp}
\end{eqnarray}
where
\begin{eqnarray}
\Pi_j^{({\rm in},k)}&=&J_m(x_k)\psi_j^{({\rm in},k)},\nonumber\\
\Pi_j^{({\rm ex},k)}&=&K_m(y_k)\psi_j^{({\rm ex},k)},\nonumber\\
\Xi_j^{({\rm in},k)}&=&\frac{J'_m(x_k)}{x_k}\psi_j^{({\rm in},k)},\nonumber\\
\Xi_j^{({\rm ex},k)}&=&\frac{K'_m(y_k)}{y_k}\psi_j^{({\rm
ex},k)},\nonumber\\
{\mathcal{D}}_j^{({\rm in},k)}&=&-i\frac{B^2
m^2\sum\limits_{l=1}^{+\infty}\phi _l^{({\rm in},k)}\Pi_l^{({\rm
in},k)}}{4R^3\Big<\phi ^{({\rm in},k)}\Big|L_{A_1}\Big|\phi^{({\rm
in},k)}\Big>}\phi_j^{({\rm in},k)},
\end{eqnarray}
and
\begin{eqnarray}
\psi_j^{({\rm in},k)} =\Big\{ {\begin{array}{*{20}c}
   (\frac{L}{\pi R})^2\frac{\tilde{\omega}^2S_{kj}}{j^2-k^2} & {k\ne j}  \\
   1 & {k=j}  \\
\end{array}},
\end{eqnarray}
\begin{eqnarray}
\psi_j^{({\rm ex},k)} =\Big\{ {\begin{array}{*{20}c}
   (\frac{\rho_{\rm ex}}{\rho_{\rm in}})(\frac{L}{\pi R})^2\frac{\tilde{\omega}^2S_{kj}}{j^2-k^2} & {k\ne j}  \\
   1 & {k=j}  \\
\end{array}},
\end{eqnarray}
$x_{k}=\left |k_{{\rm in},k}\right|R$ and $y_k=k_{{\rm ex},k}R$.
Note that the dispersion relation is then given by requiring that
the system (\ref{dp}) has non-trivial solutions i.e. its
determinant is zero.

%-----------------------------------------------------------------------------------------------
\section{Numerical results}
As typical parameters for a coronal loop, we adopt radius =
$10^{3}$ km, length = $10^{5}$ km, $\rho_{\rm in}=2\times
10^{-14}$ gr cm$^{-3}$, $\rho_{\rm ex}/\rho_{\rm in}=0.1$, $B=100$
G. For these parameters one finds $v_{A_{\rm in}}=2000$ km
s$^{-1}$, $v_{A_{\rm ex}}=6400$ km s$^{-1}$ and $\omega_{A_{\rm
in}}:=\frac{v_{A_{\rm in}}}{{\rm L}}=0.02$ rad s$^{-1}$.

\begin{figure}
\center \includegraphics{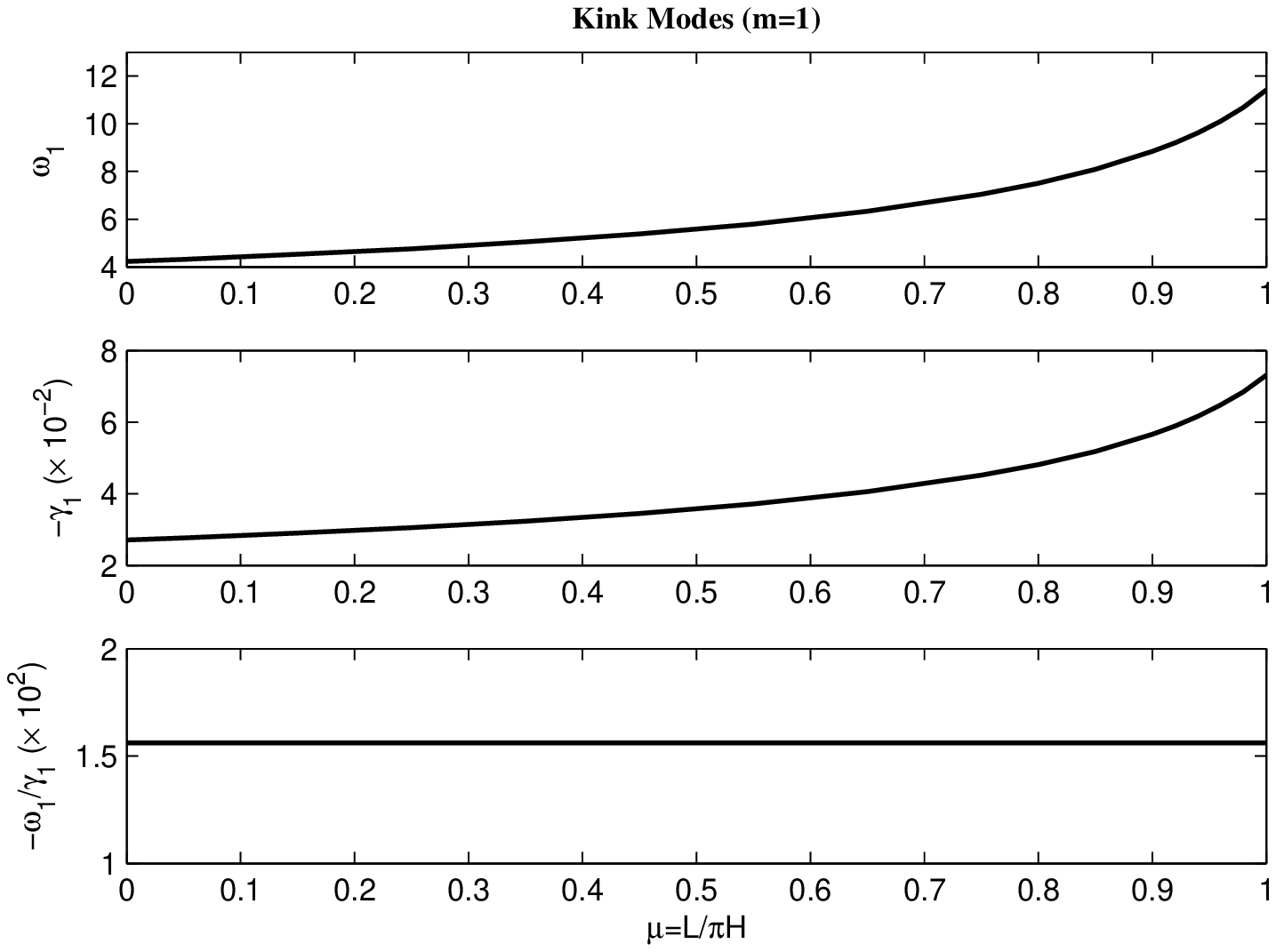}
      \vspace{9.2cm}
      \caption[]{Frequency of the fundamental kink mode ($m=1$) and its damping
rate as well as the ratio of the oscillation frequency to the
damping rate as a function of the stratification parameter
$\mu=L/\pi H$, for a loop with constant length and varying
scaleheight. The loop parameters are: $L=10^5$ km, $R/L=0.01$,
$l/R=0.02$, $\rho_{e}/\rho_{i}=0.1$, $\rho_{i}=2\times 10^{-14}$
gr cm$^{-3}$, $B=100$ G. Both frequencies and damping rates are in
units of the interior Alfv\'{e}n frequency, $\omega_{\rm
A_i}=0.02{\rm~rad~s^{-1}}$.}
         \label{m1l1-omegagamma}
   \end{figure}

\begin{figure}
\center \includegraphics{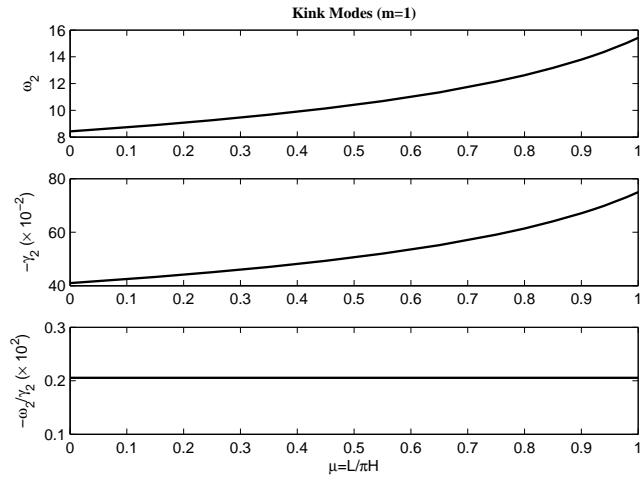}
      \vspace{9.2cm}
      \caption[]{Same as Fig. \ref{m1l1-omegagamma}, for the first-overtone kink
      modes.}
         \label{m1l2-omegagamma}
   \end{figure}

\begin{figure}
\center \includegraphics{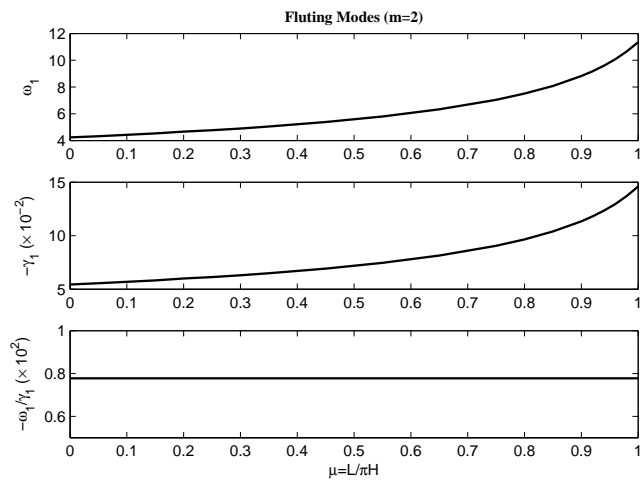}
      \vspace{9.2cm}
      \caption[]{Frequency of the fundamental fluting mode ($m=2$) and its damping
rate as well as ratio of the frequency and the damping rate as a
function of the stratification parameter $\mu=L/\pi H$. Auxiliary
parameters as in Fig. \ref{m1l1-omegagamma}.}
         \label{m2l1-omegagamma}
   \end{figure}

\begin{figure}
\center \includegraphics{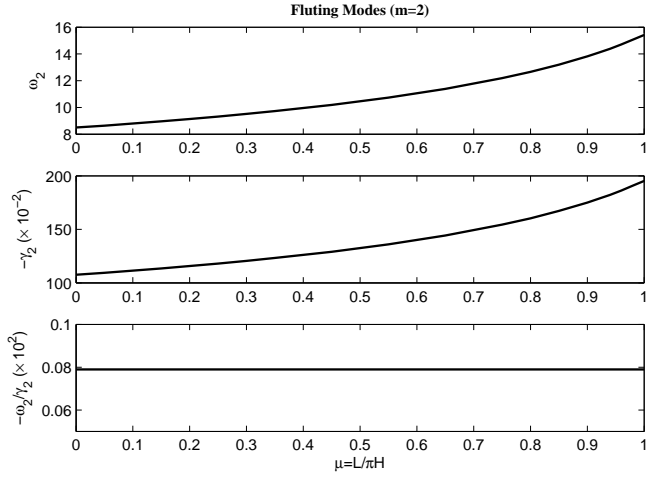}
      \vspace{9.2cm}
      \caption[]{Same as Fig. \ref{m2l1-omegagamma}, for the first-overtone fluting
      modes.}
         \label{m2l2-omegagamma}
   \end{figure}

\begin{figure}
\center \includegraphics{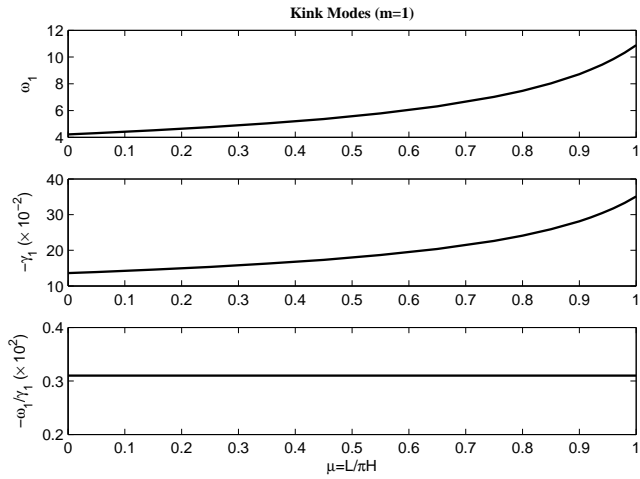}
      \vspace{9.2cm}
      \caption[]{Same as Fig. \ref{m1l1-omegagamma}, for the fundamental kink mode ($m=1$) with $l/R=0.1$.}
         \label{m1l1-omegagamma-thick}
   \end{figure}

\begin{figure}
\center \includegraphics{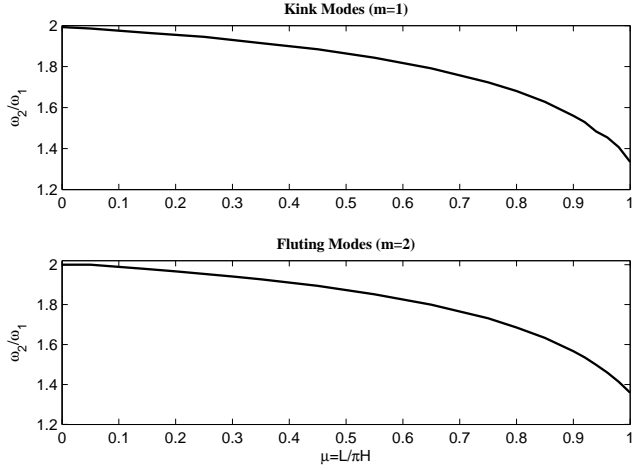}
      \vspace{9.2cm}
      \caption[]{Ratio of the frequencies $\omega_2/\omega_1$ of the first-overtone and its fundamental mode versus
      $\mu=L/\pi H$ for both kink ($m=1$) and fluting ($m=2$) modes. Auxiliary parameters as in Fig. \ref{m1l1-omegagamma}.}
         \label{m1m2-omega2omega1}
   \end{figure}

\begin{figure}
\center \includegraphics{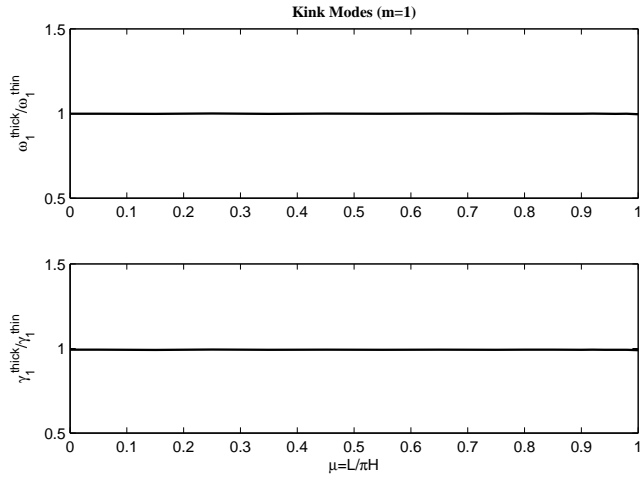}
      \vspace{9.2cm}
      \caption[]{Ratios of the frequencies $\omega_1^{\rm thick}/\omega_1^{\rm thin}$ and the damping rates $\gamma_1^{\rm thick}/\gamma_1^{\rm thin}$
      of the fundamental kink ($m=1$) mode versus $\mu=L/\pi H$ for two tube radii $R/L=$ 0.02 (thick)
and 0.01 (thin). Other auxiliary parameters as in Fig.
\ref{m1l1-omegagamma}.}
         \label{m1-RL-omegagamma}
   \end{figure}

The effects of density stratification on both the frequencies
$\omega$ and damping rates $\gamma$ are calculated by numerical
solution of the dispersion relation, Eq. (\ref{dp}). The results
are displayed in Figs. \ref{m1l1-omegagamma} to
\ref{m1-RL-omegagamma}. Figures
\ref{m1l1-omegagamma}-\ref{m2l2-omegagamma} show the frequencies
and damping rates as well as the ratio of the oscillation
frequency to the damping rate of the fundamental and
first-overtone, $k=1,2$, kink ($m=1$) and fluting $(m=2)$ modes
versus the stratification parameter $\mu=L/\pi H$ for a loop with
constant length and varying scale hight. Figures
\ref{m1l1-omegagamma} to \ref{m2l2-omegagamma} reveal that: i)
Both frequencies, $\omega_1$, $\omega_2$ and their corresponding
damping rates, $|\gamma_1|$, $|\gamma_2|$ increase when $\mu$
increases. The results for $\omega_1$ and $\gamma_1$ are in
agreement with those obtained by Andries et al. (2005b) and
Arregui et al. (2005). Also the results for all $\omega$s and
$\gamma$s are in accordance with that obtained in Paper I. ii) For
$m=1$ with increasing $\mu$, the damping rates, for instance
$|\gamma_1|$, increase ($\simeq$ 2-170 percent) compared with a
non-stratified loop. 3) The ratio of the oscillation frequency to
the damping rate, $\omega/|\gamma|$, is independent of
stratification. This result is in good concord with those obtained
numerically by Andries et al. (2005b) and analytically by Dymova
\& Ruderman (2006).

Figure \ref{m1l1-omegagamma-thick} same as Fig.
\ref{m1l1-omegagamma} shows the result of frequency, damping rate
and ratio $\omega/|\gamma|$ for the fundamental kink modes but for
a thick inhomogeneous layer with $l/R=0.1$. Comparing the results
obtained for the ratio $\omega_1/|\gamma_1|$ in Fig.
\ref{m1l1-omegagamma} and \ref{m1l1-omegagamma-thick} shows that
the values of $\omega/|\gamma|$ for $l/R=0.02$ in Fig.
\ref{m1l1-omegagamma}, five times greater then the case of
$l/R=0.1$ in Fig. \ref{m1l1-omegagamma-thick}. This result is in
good agreement with that obtained by Dymova \& Ruderman (2006) who
showed that under the assumption of uniform stratification, i.e.
$\rho_{\rm ex}(z)/\rho_{\rm in}(z)$=const., the ratio
$\omega/|\gamma|$ for kink modes in thin tube-thin inhomogeneous
layer approximation is proportional to $(l/R)^{-1}$ and is
independent of stratification.

The ratio of the frequencies $\omega_2/\omega_1$ of the
first-overtone and its fundamental mode for both the kink ($m=1$)
and fluting ($m=2$) modes are plotted in Fig.
\ref{m1m2-omega2omega1}. Figure shows that for both modes, the
ratio of the frequencies decreases from 2 (for unstratified loop)
and approaches below 1.4 with increasing density stratification.
The result of $\omega_2/\omega_1$ for $m=1$ is in accordance with
that obtained by Andries et al. (2005a) and Safari et al. (2007).
Also the results for both $m=1$ and $m=2$ are in agreement with
those obtained in Paper I. For $m=1$, $\mu=0.59$ and 0.89, the
ratios $\omega_2/\omega_1$ are 1.821 and 1.576, respectively.
These are in good agreement with the frequency ratios observed by
Van Doorsselaere et al. (2007), 1.82$\pm$0.08 and 1.58$\pm$0.06,
respectively, deduced from the observations of TRACE. Fig.
\ref{m1m2-omega2omega1} does not show a noticeable difference
between the kink and fluting modes, but our numerical values shows
that for a given stratification parameter $\mu$, the frequency
ratio $\omega_2/\omega_1$ increase slightly when the azimuthal
mode number $m$ increases.

Figure \ref{m1-RL-omegagamma} shows the ratios of frequencies
$\omega_1^{\rm thick}/\omega_1^{\rm thin}$ and damping rates
$\gamma_1^{\rm thick}/\gamma_1^{\rm thin}$
      of the fundamental, $k=1$, kink modes ($m=1$) versus the stratification
parameter for two tube radii $R/L=$ 0.02 (thick) and 0.01 (thin).
Figure \ref{m1-RL-omegagamma} clears that both the frequencies and
damping rates are independent of the tube radius, $R$, in the
limit of slender tubes (see e.g. Van Doorsselaere et al. 2004a).
%-----------------------------------------------------------------------------------------------
\section{Conclusions}
Resonant absorption of standing fast MHD body waves in coronal
loops in presence of both the longitudinal density stratification
and radial density structuring is studied. The radial structuring
is assumed to be a linearly varying density profile. To do this, a
typical coronal loop is considered as a straight pressureless
cylindrical flux tube embedded in a constant background magnetic
field. Using the relevant connection formulae, the dispersion
relation is obtained and solved numerically for obtaining both the
frequencies and damping rates of the fundamental and
first-overtone kink and fluting modes. Our numerical results show
that:

i) Both frequencies and damping rates of the fundamental and
first-overtone modes of both the kink ($m=1$) and fluting ($m=2$)
waves increase when the stratification parameter increases.

ii) The ratio of the oscillation frequency to the damping rate,
$\omega/|\gamma|$ for both the kink ($m=1$) and fluting ($m=2$)
modes is independent of stratification.

iii) The ratio of the frequencies $\omega_2/\omega_1$ for both the
kink ($m=1$) and fluting ($m=2$) modes are lower than 2 (for
unstratified loops), respectively, in presence of the longitudinal
density stratification. The result of $\omega_2/\omega_1$ for kink
modes is in accord with the TRACE observations.

Note that in our work, calculating the frequencies
$\omega_1,\omega_2$ and their corresponding damping rates
$\gamma_1,\gamma_2$ as well as the ratio of frequencies
$\omega_2/\omega_1$, for fluting modes (m=2) is a new result in
comparison with Andries et al. (2005a,b) and the rest is a
repetition of the results previously obtained by other authors.
%-----------------------------------------------------------------------------------------------
\\
\\
\noindent{{\bf Acknowledgements}}. The authors thank to Prof.
Michael Ruderman for his illuminating remarks that has enabled
them to improve the clarity of the paper. This work was supported
by the Department of Physics, University of Kurdistan, Sanandaj,
Iran; the Research Institute for Astronomy $\&$ Astrophysics of
Maragha (RIAAM), Maragha, Iran.
%-----------------------------------------------------------------------------------------------

%-----------------------------------------------------------------------------------------------
%-----------------------------------------------------------------------------------------------
\clearpage
%-----------------------------------------------------------------------------------------------
%\clearpage
%-----------------------------------------------------------------------------------------------
%\clearpage
%-----------------------------------------------------------------------------------------------
%\clearpage
%-----------------------------------------------------------------------------------------------
%\clearpage
%-----------------------------------------------------------------------------------------------
%\clearpage
%-----------------------------------------------------------------------------------------------
%\clearpage
%-----------------------------------------------------------------------------------------------

\end{document}